\documentclass[10pt, conference]{IEEEtran}
\IEEEoverridecommandlockouts
\usepackage{cite}
\usepackage{amsmath,amssymb,amsfonts}
\usepackage{algorithmic}
\usepackage{graphicx}
\usepackage{textcomp}
\usepackage{xcolor}
\usepackage{pgfplots}
\usepackage{array}
\usepackage{microtype}
\usepackage{caption}
\usepackage{subcaption}
\pgfplotsset{compat=1.14} 
\usepackage{tikz,tikzscale}
\usetikzlibrary{shapes.geometric, arrows,positioning,calc,decorations.pathreplacing}
\usepackage{standalone}
\usepackage{url}
\newcommand{\ignore}[1]{}
\def\BibTeX{{\rm B\kern-.05em{\sc i\kern-.025em b}\kern-.08em
    T\kern-.1667em\lower.7ex\hbox{E}\kern-.125emX}}

\begin{document}

\title{Secret Sharing MPC on FPGAs in the Datacenter}

\author{\IEEEauthorblockN{Pierre-Fran\c{c}ois Wolfe\IEEEauthorrefmark{1},
Rushi Patel\IEEEauthorrefmark{2}, Robert Munafo\IEEEauthorrefmark{3},
Mayank Varia\IEEEauthorrefmark{4}, and Martin Herbordt\IEEEauthorrefmark{5}}
\IEEEauthorblockA{
\IEEEauthorrefmark{1}
\IEEEauthorrefmark{2}
\IEEEauthorrefmark{3}
\IEEEauthorrefmark{5}
Dept. of Electrical and Computer Engineering \& \IEEEauthorrefmark{4}Dept. of Computer Science,\\
Boston University,
Boston, USA\\
Email: \IEEEauthorrefmark{1}pwolfe@bu.edu,
\IEEEauthorrefmark{2}ruship@bu.edu,
\IEEEauthorrefmark{3}rmunafo@bu.edu,
\IEEEauthorrefmark{4}varia@bu.edu,
\IEEEauthorrefmark{5}herbordt@bu.edu}}

\maketitle


\begin{abstract}

Multi-Party Computation (MPC) is a technique enabling data from several sources to be used in a secure computation revealing only the result while protecting the original data, facilitating shared utilization of data sets gathered by different entities.
The presence of Field Programmable Gate Array (FPGA) hardware in datacenters can provide accelerated computing as well as low latency, high bandwidth communication that bolsters the performance of MPC and lowers the barrier to using MPC for many applications.
In this work, we propose a Secret Sharing FPGA design based on the protocol described by Araki et al. \cite{araki16}.
We compare our hardware design to the original authors' software implementations of Secret Sharing and to work accelerating MPC protocols based on Garbled Circuits with FPGAs.
Our conclusion is that Secret Sharing in the datacenter is competitive and when implemented on FPGA hardware was able to use at least 10$\times$ fewer computer resources than the original work using CPUs.

\end{abstract}

\section{Introduction}

Many organizations face the problem of wanting to perform useful computations when the underlying data is sensitive.
Cryptographically secure multi-party computation (MPC) allows people to outsource encoded versions of their data to several compute parties, who can then analyze the data without reading it.
As defined in pending legislation within the United States Senate, ``the term `secure multi-party computation' means a computerized system that enables different participating entities in possession of private sets of data to link and aggregate their data sets for the exclusive purpose of performing a finite number of pre-approved computations without transferring or otherwise revealing any private data to each other or anyone else'' \cite{Wyden2019}.

MPC has been an active area of research for about 40 years \cite{evans18,Yao1982,Yao1986,Shamir1979},
and it has been deployed to protect data in the healthcare \cite{DBLP:journals/cj/ArcherBLKNPSW18,mpc-hospital}, education \cite{DBLP:journals/iacr/BogdanovKKRST15,feigenbaum2004secure}, finance \cite{DBLP:conf/fc/BogetoftCDGJKNNNPST09,DBLP:conf/fc/DamgardDNNT16,DBLP:conf/cans/AbidinACM16}, and technology \cite{DBLP:conf/ccs/BonawitzIKMMPRS17,DBLP:journals/iacr/IonKNPRSSSY19} sectors.
Nevertheless, recent surveys reveal a few companies with specialized MPC offerings 
For adoption of MPC to increase, it is necessary to continue to improve the performance and ease of use of general-purpose systems.
%
Existing work has shown that acceleration of general-purpose MPC can translate into viable systems 
\cite{DBLP:conf/ndss/HuangEK12}.
The crux of this and related work is whether MPC is amenable to hardware acceleration.
There are two main techniques for achieving MPC: Secret Sharing consumes significantly lower bandwidth but requires low latency, and Garbled Circuits are compute-bound in any environment with sufficiently high bandwidth.
Between these two options, Garbled Circuits appear more amenable to hardware acceleration, which is the subject of substantial prior research, especially with FPGAs \cite{Jarvinen2010,Jarvinen2010b,Frederiksen2014,songhori16,hussain18,songhori19,hussain19,Fang2017a,Fang2019,Huang2019a,Leeser2019}.
%
%
However, the overall trend of consolidating computing into data centers changes this calculus. Evans et al. note that ``[b]andwidth within a data center is inexpensive" with the caveat that there are security questions that must be given careful consideration in this context \cite{evans18}.
Our exploration finds a compelling argument for hardware acceleration of MPC via Secret Sharing: when deployed in a datacenter,
the low latency between accelerators (e.g., within a node, bump-in-the-wire, etc.) can enable Secret Sharing MPC to make more effective use of the available bandwidth than their Garbled Circuit counterparts \cite{Schneider2013}.
We propose FPGAs as a hardware platform to maximize the performance of MPC in the datacenter because they provide high bandwidth and minimize latency by integrating compute and communication. 

In this paper, we
explore different datacenter models,
consider the steps necessary to create a viable MPC cloud service,
examine the trade-off between Garbled Circuits and Secret Sharing, 
implement Secret Sharing in hardware,
test this hardware design, and
assess its scalability.
We conclude by proposing directions for future work
toward a complete MPC cloud service.




We summarize the contributions in this work:
\begin{itemize}
    \item We believe we are the first group to report on Secret Sharing MPC on FPGA hardware. We demonstrate that given a set of reasonable security assumptions, MPC on FPGAs in the datacenter is viable for a real service. 
    \item We demonstrate that Secret Sharing outperforms state-of-the-art methods for implementing MPC in the datacenter.
    \item Using 5.5\% of FPGA fabric in a consumer cloud environment, we match the throughput of an optimized 20-core CPU implementation saturating a typical 10Gbps network connection. This result scales with available bandwidth: a single FPGA is able to saturate a 200Gbs link with a throughput of $\sim$26 million AES operations per second.
\end{itemize}

\section{Background}
\label{sec:background}


\subsection{Datacenter Model} 

The primary motivation of this work is to create an effective cloud datacenter that can offer MPC-as-a-service that is easy to use and has high performance.
%
%
%
Because MPC requires multiple computing parties for security and low latency networking for performance, we consider processing hardware owned by different parties and housed within a single datacenter.
This arrangement permits secure data storage across the computing parties close to processing locations.
Concretely, we imagine a scenario where a small number of FPGAs are connected over high-speed interconnects and have the benefit of drawing data from servers all co-located within the datacenter.

FPGA hardware acceleration has seen increasing adoption in datacenters.
As described in Section \ref{sub:mpc-design}, FPGA hardware properties and co-location yield high throughput for MPC protocols based on Secret Sharing, which makes the most effective use of available bandwidth.
Maximizing throughput is a focus for this work as this metric determines how efficiently multiple client tasks can be completed.
%
%

\subsection{MPC Paradigms}
\label{sub:mpc-design}


MPC protocols support an arbitrary number $N$ of compute parties and tolerate an arbitrary threshold $T$ of `bad' parties working together, where this coalition might try to break confidentiality to learn other people's data or to tamper with the integrity of the calculation.
In this work, we examine a 3-party protocol tolerating 1 adversarial party who ``semi-honestly'' follows the protocol and only tries to break confidentiality. This matches a scenario in which a small number of FPGAs owned by different parties are co-located within a datacenter.

General-purpose MPC designs often represent the agreed-upon computation as an arithmetic or Boolean circuit, and follow the Garbled Circuit or Secret Sharing approaches.
Garbled Circuits rely on one compute party generating a (large) encoded version of the entire circuit, which it then transmits to a second party who can evaluate the encoded circuit on encoded inputs in order to recover the answer.
On the other hand, Secret Sharing-based MPC systems have the compute parties evaluate each gate of the circuit in parallel on their own pieces or \emph{shares} of the data, with a small amount of network communication required for each multiplication or AND gate (none is required for addition or XOR gates).

The computation and communication overhead of MPC manifests itself differently for Garbled Circuits and Secret Sharing.
Even with optimizations \cite{Beaver1990,Naor1999,Kolesnikov2008,DBLP:conf/eurocrypt/ZahurRE15,Yakoubov2017}, Garbled Circuits have a small number of communication rounds but a large communication size (80-128$\times$ the size of the original data), rendering them beneficial in high-latency scenarios but detrimental when processing large datasets.
Conversely, Secret Sharing approaches require a low-latency environment because they involve many rounds of communication, however they consume substantially less bandwidth per computational step.


To date, most MPC implementations are in software, and thus rely on general-purpose processing hardware and commodity networking equipment.
In this scenario, Secret Sharing tends to be network latency-bound whereas Garbled Circuits are often compute-bound. Consequently, most of the prior focus in hardware acceleration has been directed toward Garbled Circuits.
Our work specifically considers MPC implementations in the datacenter, where Secret Sharing systems offer higher maximum throughput and the network latency can be low enough to realize meaningful performance benefits by optimizing the computation with FPGAs.
\tikzstyle{startstop} = [rectangle, rounded corners, minimum width=1cm, minimum height=0.5cm,text centered, draw=black, fill=red!30]
\tikzstyle{io} = [trapezium, trapezium left angle=70, trapezium right angle=110, minimum width=1cm, minimum height=0.5cm, text centered, draw=black, fill=blue!30]
\tikzstyle{process} = [rectangle, minimum width=1cm, minimum height=0.5cm, text centered, draw=black, fill=orange!30]
\tikzstyle{data} = [rectangle, minimum width=1cm, minimum height=0.5cm, text centered, draw=black, fill=green!30]
\tikzstyle{arrow} = [thick,->,>=stealth]
\tikzstyle{every node}=[font=\footnotesize] 




\subsection{Selected MPC Protocol}
\label{sec:application}


Within the category of MPC protocols based on Secret Sharing, we selected a protocol by Araki et al.\ \cite{araki16,araki16b} for FPGA acceleration due to its simplicity and its impressive performance in software.
The Araki et al.\ protocol employs exactly 3 parties, and it tolerates 1 adversarial party that is presumed to follow the protocol.
Also, communication occurs in a ring topology, where each party only needs to communicate with 1 of the other 2 parties.

The workflow involves 3 distinct steps. First, data holders split their data into \emph{shares} held by the 3 compute parties. Then, the parties iteratively \emph{compute} over these shares without revealing any secrets. Finally, the compute parties reveal their shares to the output party who can \emph{reconstruct} the final answer.

\begin{figure}
\begin{center}
  \begin{tikzpicture}[scale=1, node distance=0.4cm]
    \node (start) [startstop] {Start Secret Share};
    \node (rand1) [process, below = of start] {Random};
    \node (in1) [io, left = of rand1] {Value};
    \node (rand2) [process, right = of rand1] {Random};
    \node (xor1) [process, below = of $(rand1)!0.5!(rand2)$] {XOR};
    \node (x3) [data, below = of xor1] {$X_3$};
    \node (x1) [data, left = of x3] {$X_1$};
    \node (x2) [data, right = of x3] {$X_2$};
    \node (xor2) [process, below = of x1] {XOR};
    \node (xor3) [process, below = of x2] {XOR};
    \node (xor4) [process, below = of x3] {XOR};
    \node (a1) [data, below = of xor4] {$A_1$};
    \node (a2) [data, below = of xor2] {$A_2$};
    \node (a3) [data, below = of xor3] {$A_3$};
    \node (s3) [io, below = of a3] {$X_3A_3$};
    \node (s1) [io, left = of s3] {$X_1A_1$};
    \node (s2) [io, left = of s1] {$X_2A_2$};
    \node (stop) [startstop, below = of s1] {Finish Secret Share};
    \draw [arrow] (start.south) |- ($(start)!0.5!(rand1)$) -| (in1.north);
    \draw [arrow] (start.south) |- ($(start)!0.5!(rand1)$)  -| (rand1.north);
    \draw [arrow] (start.south) |- ($(start)!0.5!(rand1)$) -| (rand2.north);
    \draw [arrow] (in1.south) |- (xor2.west);
    \draw [arrow] (in1.south) |- ($(x1)!0.5!(xor4)$) |- (xor4.west);
    \draw [arrow] (in1.south) |- ($(x3)!0.5!(xor3)$) |- (xor3.west);
    \draw [arrow] (rand1.south) |- (xor1.west);
    \draw [arrow] (rand2.south) |- (xor1.east);
    \draw [arrow] let \p1 = (rand1), \p2 = (xor1) in (rand1.south) |- (\x1,\y2) -| (x1.north);
    \draw [arrow] let \p1 = (rand2), \p2 = (xor1) in (rand2.south) |- (\x1,\y2) -| (x2.north);
    \draw [arrow] (xor1.south) -- (x3.north);
    \draw [arrow,red] (x1.south) -- (xor2.north);
    \draw [arrow,green] (x2.south) -- (xor3.north);
    \draw [arrow,blue] (x3.south) -- (xor4.north);
    \draw [arrow,red] let \p1 = ($(a2)!0.45!(a1)$), \p2 = ($(a1)!0.5!(s1)$), \p3 = ($(s2)!0.5!(s1)$) in (x1.east) -| (\x1,\y2)  |- (\x3,\y2) |- (s1.west);
    \draw [arrow,green] let \p1 = ($(a1)!0.5!(a3)-(a1)+(a3)$), \p2 = ($(a2)!0.5!(xor2)$), \p3 = ($(s2)!0.5!(s1)-(s1)+(s2)$) in (x2.east) -| (\x1,\y2)  -| (\x3,\y2) |- (s2.west);
    \draw [arrow,blue] let \p1 = ($(a1)!0.45!(a3)$), \p2 = ($(a3)!0.5!(s3)$), \p3 = ($(s1)!0.5!(s3)$) in (x3.east) -| (\x1,\y2)  |- (\x3,\y2) |- (s3.west);
    \draw [arrow] (xor2.south) -- (a2.north);
    \draw [arrow] (xor3.south) -- (a3.north);
    \draw [arrow] (xor4.south) -- (a1.north);
    \draw [arrow] (a1.south) |- ($(a1)!0.5!(s1)$) -| (s1.north);
    \draw [arrow] (a2.south) |- ($(a2)!0.5!(s2)$) -| (s2.north);
    \draw [arrow] (a3.south) |- ($(a3)!0.5!(s3)$) -| (s3.north);
    \draw [arrow] (s1.south) |- ($(s1)!0.5!(stop)$) -| (stop.north);
    \draw [arrow] (s2.south) |- ($(s1)!0.5!(stop)$) -| (stop.north);
    \draw [arrow] (s3.south) |- ($(s1)!0.5!(stop)$) -| (stop.north);
\end{tikzpicture}
\end{center}
\vspace*{-0.1truein}
\caption{Initial Secret Sharing}
\label{fig:share}
\vspace*{-0.2truein}
\end{figure}
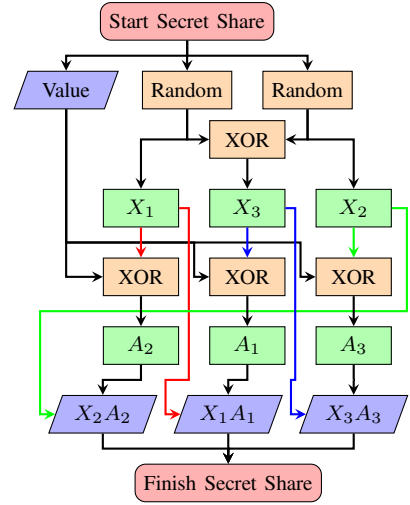

In the \emph{share phase}, anyone holding a secret value $v \in \{0,1\}$ can split this secret among the 3 compute parties as follows.
\begin{itemize}
    \item The data holder selects $x_1, x_2, x_3$ uniformly from $\{0,1\}$
    subject to the constraint that $x_1 \oplus x_2 \oplus x_3 = 0$.
    \item The data holder sends 
    $(x_i, a_i$) to
    each compute party, where
    $a_i = x_{i-1} \oplus v$ is a one-time pad of the secret.
\end{itemize}
The one-time pad hides the secret value $v$ from any single party $P_i$.
See Figure \ref{fig:share} for an illustration.

\begin{figure}
\begin{center}
    \begin{tikzpicture}[node distance=0.4cm]
    \node (start) [startstop] {Start XOR Computation};
    \node (s1) [io, below left = of start] {$X_iA_i$};
    \node (s2) [io, below right = of start] {$Y_iB_i$};
    \node (yi) [data, below right = 0.5cm of s1] {$Y_i$};
    \node (xi) [data, below left = 0.5cm of s1] {$X_i$};
    \node (ai) [data, below left = 0.5cm of s2] {$A_i$};
    \node (bi) [data, below right = 0.5cm of s2] {$B_i$};
    \node (xor1) [process, below = 0.3cm of yi] {XOR};
    \node (xor2) [process, below = 0.3cm of ai] {XOR};
    \node (zi) [data, below = 0.3cm of xor1] {$Z_i$};
    \node (ci) [data, below = 0.3cm of xor2] {$C_i$};
    \node (s3) [io, below = of $(zi)!0.5!(ci)$)] {$Z_iC_i$};
    \node (stop) [startstop, below = 0.3cm of s3] {Finish XOR Computation};
    \draw [arrow] (start.south) |- ($(start.south) - (0,1mm)$) -| (s1.north);
    \draw [arrow] (start.south) |- ($(start.south) - (0,1mm)$) -| (s2.north);
    \draw [arrow,red] (s1.south) |- ($(s1.south) - (0,1mm)$) -| (xi.north);
    \draw [arrow,red] (s1.south) |- ($(s1.south) - (0,1mm)$) -| (ai.north);
    \draw [arrow,blue] (s2.south) |- ($(s2.south) - (0,1.5mm)$) -| (yi.north);
    \draw [arrow,blue] (s2.south) |- ($(s2.south) - (0,1.5mm)$) -| (bi.north);
    \draw [arrow] (xi.south) |- (xor1.west);
    \draw [arrow] (ai.south) -- (xor2.north);
    \draw [arrow] (yi.south) -- (xor1.north);
    \draw [arrow] (bi.south) |- (xor2.east);
    \draw [arrow] (xor1.south) -- (zi.north);
    \draw [arrow] (xor2.south) -- (ci.north);
    \draw [arrow] (zi.south) |- (s3.west);
    \draw [arrow] (ci.south) |- (s3.east);
    \draw [arrow] (s3.south) -- (stop.north);
\end{tikzpicture}
\end{center}
\vspace*{-0.1truein}
\caption{Party i's contribution toward computing an XOR gate}
\label{fig:xor}
\vspace*{-0.1truein}
\end{figure}
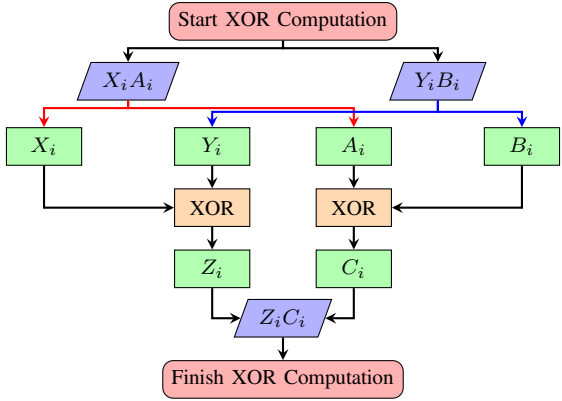

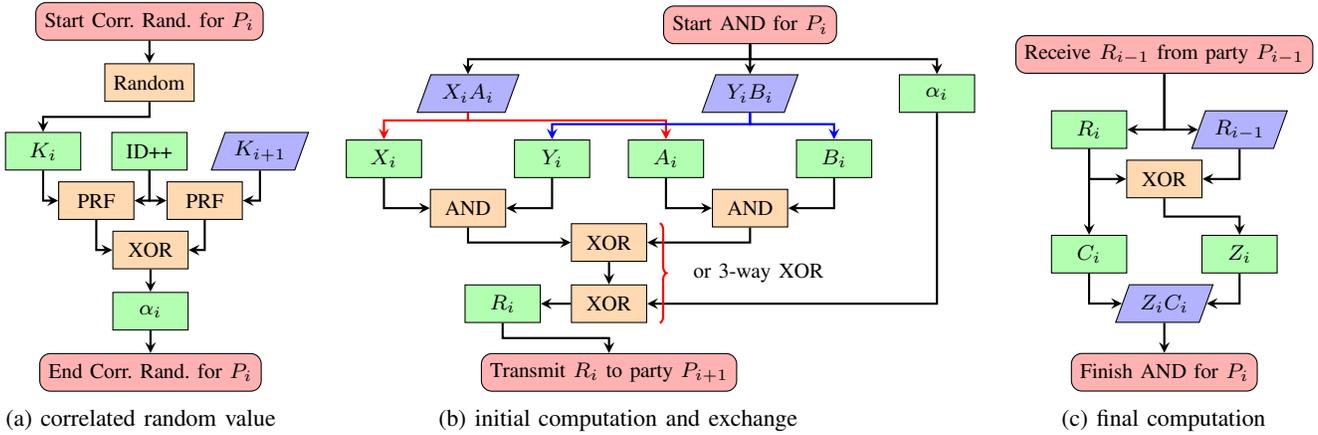
\begin{figure*}
    \centering
    \begin{subfigure}[b]{0.2\textwidth}%
        \centering
        \begin{tikzpicture}[node distance=0.4cm]
    \node (P1)  [startstop] {Start Corr. Rand. for $P_i$};
    \node (Rand1) [process, below = 0.3cm of P1] {Random};
    \node (ID1) [data, below = of Rand1] {ID++};
    \node (K1) [data, left = of ID1] {$K_i$};
    \node (K2) [io, right = of ID1] {$K_{i+1}$};
    \node (A1) [process, below = of $(K1)!0.5!(ID1)$] {PRF};
    \node (A2) [process, below = of $(ID1)!0.5!(K2)$] {PRF};
    \node (X1) [process, below = of $(A1)!0.5!(A2)$] {XOR};
    \node (alpha) [data, below = 0.3cm of X1] {$\alpha_i$};
    \node (end) [startstop, below = 0.3cm of alpha] {End Corr. Rand. for $P_i$};
    \draw [arrow] (P1.south) -- (Rand1.north);
    \draw [arrow] (K1.south) |- (A1.west);
    \draw [arrow] (Rand1.south) |- ($(Rand1)!0.5!(ID1)$) -| (K1.north);
    \draw [arrow] (ID1.south) |-  (A1.east);
    \draw [arrow] (ID1.south) |- (A2.west);
    \draw [arrow] (K2.south) |- (A2.east);
    \draw [arrow] ($(A1.south)$)  |- (X1.west);
    \draw [arrow] ($(A2.south)$) |- (X1.east);
    \draw [arrow] (X1.south) -- (alpha.north);
    \draw [arrow] (alpha.south) -- (end.north);
 \end{tikzpicture}
        \caption{correlated random value}
        \label{fig:and_1}
    \end{subfigure}%
    \hfill
    \begin{subfigure}[b]{0.4\textwidth}%
        \centering
        \begin{tikzpicture}[node distance=0.4cm]
    \node (start) [startstop] {Start AND for $P_i$};
    \node (Y1B1)  [io, below = of start] {$Y_iB_i$};
    \node (A1)    [data, below left = 0.5cm of Y1B1] {$A_i$};
    \node (B1)    [data, below right = 0.5cm of Y1B1] {$B_i$};
    \node (Alpha) [data, above right = 0.5cm of B1] {$\alpha_i$};
    \node (Y1)    [data, left = 0.5cm of A1] {$Y_i$};
    \node (X1A1)  [io, above left = 0.5cm of Y1] {$X_iA_i$};
    \node (X1)    [data, below left = 0.5cm of X1A1] {$X_i$};
    \node (AND1)  [process, below = of $(X1)!0.5!(Y1)$ ] {AND};
    \node (AND2)  [process, below = of $(A1)!0.5!(B1)$ ] {AND};
    \node (XOR1)  [process, below = 0.2cm of $(AND1)!0.5!(AND2)$ ] {XOR};
    \node (XOR2)  [process, below = 0.3cm of XOR1 ] {XOR};
    \node (R1)    [data, left = of XOR2] {$R_i$};
    \node (stop)  [startstop, below = of XOR2] {Transmit $R_i$ to party $P_{i+1}$};
    \node (WAY3)  [right ={1cm} of $(XOR1)!0.5!(XOR2)$] {or 3-way XOR};
    \draw [arrow] (start.south) |- ($(start.south) - (0,2mm)$) -| (X1A1.north);
    \draw [arrow] (start.south) |- ($(start.south) - (0,2mm)$) -| (Y1B1.north);
    \draw [arrow] (start.south) |- ($(start.south) - (0,2mm)$) -| (Alpha.north);
    \draw [arrow,red] (X1A1.south) |- ($(X1A1.south) - (0,1mm)$) -| (X1.north);
    \draw [arrow,red] (X1A1.south) |- ($(X1A1.south) - (0,1mm)$) -| (A1.north);
    \draw [arrow,blue] (Y1B1.south) |- ($(Y1B1.south) - (0,1.5mm)$) -| (Y1.north);
    \draw [arrow,blue] (Y1B1.south) |- ($(Y1B1.south) - (0,1.5mm)$) -| (B1.north);
    \draw [arrow] (X1.south) |- (AND1.west);
    \draw [arrow] (Y1.south) |- (AND1.east);
    \draw [arrow] (A1.south) |- (AND2.west);
    \draw [arrow] (B1.south) |- (AND2.east);
    \draw [arrow] (AND1.south) |- (XOR1.west);
    \draw [arrow] (AND2.south) |- (XOR1.east);
    \draw [arrow] (XOR1.south) -- (XOR2.north);
    \draw [arrow] (Alpha.south) |- (XOR2.east);
    \draw [arrow] (XOR2.west) -- (R1.east);
    \draw [arrow] (R1.south) |- ($(XOR2)!0.5!(stop)$) -| (stop.north);
    \draw [decoration={brace,raise=5pt},decorate,thick,red] ($(XOR1.north)+(right:0.5cm)$) -- ($(XOR2.south)+(right:0.5cm)$);
\end{tikzpicture}
        \caption{initial computation and exchange}
        \label{fig:and_2}
    \end{subfigure}%
    \hfill
    \begin{subfigure}[b]{0.3\textwidth}%
        \centering
        \begin{tikzpicture}[node distance=0.4cm]
    \node (P1)  [startstop] {Receive $R_{i-1}$ from party $P_{i-1}$};
    \node (R1)  [on grid, data, below left = 1cm and 1cm of P1] {$R_i$};
    \node (R3)  [on grid, io, below right = 1cm and 1cm of P1] {$R_{i-1}$};
    \node (X1)  [process, below = of $(R1)!0.5!(R3)$] {XOR};
    \node (C1)  [on grid, data, below left = 1cm and 1cm of X1] {$C_i$};
    \node (Z1)  [on grid, data, below right = 1cm and 1cm of X1] {$Z_i$};
    \node (Z1C1)[io, below = of $(C1)!0.5!(Z1)$] {$Z_iC_i$};
    \node (end) [startstop, below = of Z1C1] {Finish AND for $P_i$};
    \draw [arrow] (P1) |- (R1.east);
    \draw [arrow] (P1) |- (R3.west);
    \draw [arrow] (R1.south) -- (C1.north);
    \draw [arrow] (R1.south) |- (X1.west);
    \draw [arrow] (R3.south) |- (X1.east);
    \draw [arrow] (X1.south) |- ($(X1)!0.5!(Z1)$) -| (Z1.north);
    \draw [arrow] (C1.south) |- (Z1C1.west);
    \draw [arrow] (Z1.south) |- (Z1C1.east);
    \draw [arrow] (Z1C1.south) -- (end.north);
\end{tikzpicture}
        \caption{final computation}
        \label{fig:and_3}
    \end{subfigure}%
    \caption{Party i's contribution toward computing an AND gate}
    \label{fig:and}
    \vspace*{-0.1truein}
\end{figure*}

In the \emph{compute phase}, the parties work together to compute shares of the result of each XOR or AND gate in a privacy-preserving manner.
It is easiest to imagine fan-in 2 operations proceeding sequentially with inputs \((x_i,a_i)\) and \((y_i,b_i)\),
though we stress that this process is embarrassingly parallel.
\begin{itemize}
    \item \emph{XOR operation}: Each share of the XOR of the two values simply equals the XOR of the two input shares
    (see Figure \ref{fig:xor})
    because a one-time pad is homomorphic under the $\oplus$ operation. The parties do not need to communicate.
    \item \emph{AND operation}:
Calculating shares of the result of an AND gate is more complex;
it requires each compute party to compute a non-trivial amount of Boolean logic and transmit one bit of information to one other compute party.
First, the parties produce \emph{correlated random values} 
$\alpha_1$, $\alpha_2$, and $\alpha_3$ that XOR to 0 but are independent of any secret values;
if the parties distribute short keys before the computation, they can generate correlated randomness quickly using a PRF such as AES (Figure \ref{fig:and_1}).
Second, each compute party follows the circuit shown in Figure \ref{fig:and_2} that consumes the correlated randomness generated above; Araki et al.\ show that the resulting values $R_i$ have the property that $R_1 \oplus R_2 \oplus R_3$ equals the result of the AND gate.
Finally, each party $P_i$ transmits $R_i$ to another party $P_{i+1}$, and then re-builds shares of the result in our desired format $\{(z_i, c_i)\}$ (Figure \ref{fig:and_3}).
    
\end{itemize}

In the \emph{reconstruction} phase, we presume that the compute parties have calculated shares $\{(x_i',a_i')\}$ corresponding to the output value $v'$. Then, the parties can reconstruct $v'$ by revealing their shares and computing $\bigoplus_{i=1}^{3} a_i' = \bigoplus_{i=1}^{3} (x_i' \oplus v) = v$.

%


There exist extensions of the Araki et al.\ protocol that permit additional parties or provide stronger security against a malicious attacker \cite{araki17,furukawa17,furukawa19}.
An FPGA implementation of this protocol provides an ideal starting point from which to explore the benefits of
acceleration for related schemes with different features, and
the possibility of
dynamically switching between them to improve
performance further
\cite{DBLP:conf/ndss/Demmler0Z15,DBLP:conf/ccs/MohasselR18}.



\subsection{FPGA Models}
Several FPGA deployment models are possible with varying trade-offs. Options from lowest to highest performance include: (1) co-processor, (2) bump-in-the-wire, (3) single-node cluster, (4) enclave/silo on FPGA. 
The enclave/silo approach where one FPGA is allocated into several regions for different parties is appealing from a performance perspective as it would enable near zero latency and near infinite bandwidth. Such an arrangement does raise many difficult questions about the isolation of the parties which go beyond the scope of the current work.
Within this hierarchy, the Amazon AWS F1 instances we consider fall into the single-node cluster category. Amazon describes two different inter-board communication approaches.The F1.4xlarge, and F1.16xlarge instances should have a 400Gbps serial ring link but support is only planned in a future release. Communication between FPGAs is possible at 12Gbps over PCIe. 
In testing the proposed Secret Sharing block, an AWS F1.2xlarge instance was used rather than the the 4x or 16x as initial testing only required a single FPGA. 


\newcolumntype{P}[1]{>{\centering\arraybackslash}p{#1}} 
\begin{table}[t]
\centering
\caption{Araki et al. Result Analysis}
\begin{tabular}{||P{0.4cm} r c P{1.3cm} c||}
\hline
\multicolumn{2}{||P{3cm}|}{Araki et al. Results} & \multicolumn{3}{|c||}{Verification} \\
\hline
Cores & \multicolumn{1}{c}{AES/sec} & Gbps/serv. & Gbps/serv. w/over. & Error \\
\hline\hline
1 & \(100103 \pm 1632\) & 0.572 & 0.559 & 2.19\% \\
\hline
5 & \(530408 \pm 7219\) & 2.99 & 2.96 & 0.85\% \\
\hline 
10 & \(975237 \pm 3049\) & 5.47 & 5.45 & 0.35\% \\
\hline
16 & \(1242310 \pm 4154\) & 6.95 & 6.94 & 0.10\% \\
\hline
20 & \(1324117 \pm 3721\) & 7.38 & 7.40 & 0.28\% \\
\hline
\end{tabular}
\label{fig:netCheck}
\vspace*{-0.1truein}
\end{table}

\section{System Design and Implementation}
\label{sec:system}
\subsection{Analysis of Original Implementation}

Obtaining the secure operation metrics enables comparison of the FPGA design to the original results.
Inspection of the original results of this Secret Sharing implementation for secure AES \cite{araki17} reveals the use of the Bristol Fashion Key Expanded AES \cite{Bristol} requiring 5440 secure AND operations. The test described by Araki et al. is embarrassingly parallel, simultaneously running 12800 independent secure AES computations per core in each node. The total AES operations performed can be used to verify the number of bits communicated. Runtime is obtained from the AES/sec rate and total number of AES operations. Including a reasonable overhead for TCP/IP of 2.74\%\cite{Iveson2013} the verified network rates closely match the reported results with less than 2.5\% error (Table \ref{fig:netCheck}).
The FPGA design can be reasonably compared to this system using the 5440 AND/AES conversion.

\subsection{FPGA Implementation}

Here we cover some FPGA implementation details for the chosen MPC protocol.
Two OpenCores projects were used, one for AES \cite{Homer2012} and one for a RNG \cite{Castillo2004} for faster development. Amazon Web Services (AWS) reference designs \cite{AWS2016} and hardware were used for testing the preliminary scheme.

At startup, each party generates a random key for the PRF and shares it with one other party.
Currently, each party uses one RNG module \cite{Castillo2004} to generate the key.
The security of this RNG block was not examined; a deployed version might use a physically unclonable function (PUF) or other secure hardware RNG.

Each party contains one PRF instance that is alternately evaluated in counter mode, using each of the two keys the party holds and the same counter. Each output pair is then XORed to produce a new correlated random number.
As the keys are only set at startup, the 21 clock cycle pipeline delay for the selected PRF was only experienced at initialization.


The MPC AND module itself consists of a few bitwise operations that produce the intermediate $R_i$ values (Figure \ref{fig:and_2}). Most latency occurs in the transmission of the $R_i$ values, since the final step (Figure \ref{fig:and_3}) cannot begin without those values.

\subsection{Analysis of FPGA Implementation}

The MPC AND hardware operation must be fed data and triggered by external logic. The first implementation uses an Arria 10 and NIOS II softcore. A NIOS custom instruction was used to load data and start the operation. The custom instruction enables simple software control of the hardware MPC AND operation. The second implementation uses Amazon Web Service (AWS) FPGAs available through its Elastic Compute Cloud (Amazon EC2). Specifically, Xilinx Virtex UltraScale+ VU9P FPGAs are accessible via a virtual machine in EC2 F1 instances. Amazon includes a hardware shell for software/hardware co-design between the node CPU (Intel Xeon E5-2686 v4) and FPGA. Software to control the FPGA uses provided DMA functions and PCIe function templates. This furnishes the mechanism for loading data, controlling operations, and retrieving results.

The PCIe packets are translated through the Amazon shell and utilize multiple AXI bus configurations to send and receive data with the software system. We use the general purpose AXI bus supporting a 512 bit data packet to provide a single message containing two secret share vectors (4$\times$128-bits) prior to starting the hardware operations. The HDL design takes each AXI bus message, parses the information, and relays data to the desired AND module. 




\subsection{Design Improvements and Additions}

Based on the minimal data dependencies and flow in Figure \ref{fig:and}, in principle a fully pipelined MPC AND module (128 Boolean MPC ANDs) can execute one AND operation per clock cycle.
Such a design would saturate a 10Gbps network connection when operated at 78.13 MHz. Operating at the higher frequencies used commonly by FPGAs would require higher bandwidth.
For example, at 200 MHz, a single MPC module of this type saturates a 25.6Gbps link. 



%

\section{Results}
\label{sec:results}

\subsection{Testing and Data}



The MPC AND module on FPGA was assessed in regards to its resource utilization with varying levels of duplication and based on the latency of evaluation. 



Initial testing targeted Intel FPGAs, such as the Arria 10, with the the NIOS II softcore executing test software to load data and trigger the MPC AND hardware.
The limitations of the Avalon Bus width and the latency a simple softcore design imposes encouraged us to consider other options.
Note, there are still circumstances where a softcore is viable such as in a design using local storage and other techniques to overcome the limits of loading individual data and running an operation in sequence.
Regardless, with the single AND design synthesized in Quartus, targeting an Arria 10 (10AX115S2F45I1SG) provided initial insights.
The most constrained resource for a single AND was the 704.5Kb of M20K block memory consumed post-synthesis, making is possible to estimate the utilization to be $\sim$1.32\% based on the total 53.260Mb of M20K available on the Arria 10.
With perfect utilization this would permit $\sim$76 instances.
More realistically perhaps 70\% of the fabric might be used allowing for $\sim$54 instances of the MPC AND.
As implemented, the MPC AND requires 6 clock cycles between operation which means that 48 MPC AND instances makes it possible for 8 operations to occur each cycle.
Even with only 8 AND operations per cycle at 200Mhz the design is able to saturate $\sim$205Gbps, far more than the 10Gbps link in the original paper. 



With these synthesis results from Quartus but seeking to avoid the limitations of the NIOS II and to find a more fitting cloud target we looked to Amazon Web Services (AWS).
While the Amazon AWS F1 instances do not currently offer the promised high-speed serial ring\cite{AWS2016}, targeting the boards available provides hardware utilization insights, and leaves the possibility of more easily using higher speed communication when support materializes.
Furthermore, the PCIe option, while lower performance, remains available for future tests.

\begin{figure}
\begin{tikzpicture}
\begin{axis}[
    xlabel={AND Cores},
    ylabel={Percent Utilization},
    xmin=0, xmax=60,
    ymin=0, ymax=100,
    xtick={0,10,20,30,40,50,60},
    ytick={0,20,40,60,80,100},
    ymajorgrids=true,
    grid style=dashed,
    width=8.5cm,
    height=5cm,
]

\addplot[
    color=blue,
    mark=square,
    ]
    coordinates {
    (0,3.2)(3,5.53)(12,14.36)(24,41.26)(48,85.6)(60,98.53)
    };
    
\end{axis}
\end{tikzpicture}
\caption{AWS FPGA fabric total utilization}
\label{fig:AWSUtilization}
\vspace*{-0.1truein}
\end{figure}
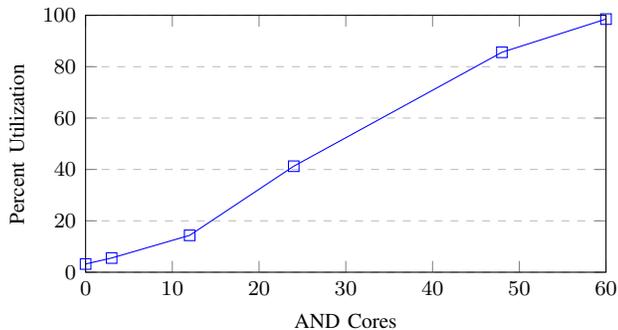

For a single 1-party block post-routing, the Virtex Ultrascale+ utilizes $\sim$3.20\% of its resources.
In order to verify that all three parties functioned together properly, a 3$\times$party design with 1 AND per party was made to target a single FPGA.
This made it possible to pass data and trigger operations without having to immediately spend the development time to bring-up the AXI4-Stream between FPGAs over PCIe.
Furthermore, since each 1 party block contains more control logic than just a single MPC AND it serves as an adequate conservative estimate of how many AND blocks might fit on one FPGA.
Continuing this line of testing led to duplicating groups of the 3$\times$party block the results of which are summarized in Table \ref{fig:AWSResults}.
The number of AND modules are used to determine the number of bits per clock cycle that are processed.
The F1 instance was clocked at 125Mhz producing the listed Gbps results. The equivalent number of AES/sec was determined by dividing by the 5440 AND/AES used in Araki et al.
A plot of the FPGA utilization in Figure \ref{fig:AWSUtilization} shows a fairly linear relationship between number of AND modules and utilization.


\begin{table}
\centering
\caption{AWS Implementation Result Analysis}
\begin{tabular}{||c c c c||}
\hline
AND Cores & Bits & Gbps & AES (millions op.)/sec \\
\hline\hline
1 & 128 & 2.67 & 0.490 \\
\hline
3 & 384 & 8.00 & 1.47 \\
\hline 
12 & 1536 & 32.0 & 5.89 \\
\hline 
24 & 3072 & 64.0 & 11.8\\
\hline
48 & 6144 & 128 & 23.5 \\
\hline
60 & 7680 & 160 & 29.4 \\
\hline
\end{tabular}
\label{fig:AWSResults}
\end{table}

\subsection{Analysis}


The tests performed with different quantities of MPC AND blocks on Amazon AWS provide sufficient data points to establish that Secret Sharing MPC can be competitive when implemented on FPGA hardware in the datacenter.
The fabric utilization in the design scales relatively linearly with increases in the quantity of AND cores.

The original work utilized a software implementation of the protocol executed on general purpose processors \cite{araki16,araki16b}. Specifically, each party used varying numbers of cores from one or two Xeon processors. The authors were able to nearly saturate their 10Gbps link (7.38Gbps) between parties when using all cores in each node. This required about 50\% of the processor's time.
While the authors were limited by multiple cores causing queuing congestion at the Network Interface Card (NIC) the use of a CPU appears to have more limited scaling potential.
Using the reported number of AES/sec and network communication for 1 core, scaling from 73.3\% CPU usage to 100\% would appear to show 1 core being capable of at most $\sim$130 thousand AES/sec, saturating a $\sim$0.780Gbps connection.
Multiplying for 20 cores that would amount to a peak of $\sim$2.7 million AES/sec and $\sim$15.6Gbps.

In comparison, the FPGA AND block we tested for performing Secret Sharing only requires 3 AND cores per party to exceed the 7.38Gbps reached with 20 CPU cores, instead being capable of 8.00Gbps. This uses $\sim$5\% of the fabric available on the FPGA targeted, a $10\times$ improvement vs the CPU utilization.

Attempting to fully employ the available fabric it is possible to implement 60 AND cores which would permit saturation of 160Gbps links while performing 29.4 million AES/sec.

These results demonstrate preferable scaling properties supporting the selection of FPGAs for acceleration.
Based on the results, targeting the anticipated 200Gbps links in the Amazon F1 would require less than 25\% fabric utilization to reach full saturation if just the pipeline improvement is made.
With a frequency improvement, even less fabric would be required.
The remaining available fabric is beneficial as it allows for work distribution and additional secure computations.

\section{Related Work}

There exists earlier research exploring hardware accelerated MPC, but the efforts have focused on Garbled Circuits rather than Secret Sharing. The hardware considered has included GPUs \cite{pu2011,Pu2013,Husted2013,Frederiksen2014}; most efforts, however, employ FPGAs.
The earliest of these efforts dates to 2010 \cite{Jarvinen2010,Jarvinen2010b}, 
with more work recently \cite{Fang2017a,hussain19,Fang2019}, and some considering Amazon AWS \cite{Leeser2019,Fang2019}.
Other work explored garbling entire processors \cite{songhori16,songhori19} and 
specialized problem acceleration \cite{hussain18}.


The studies from researchers at Northeastern University are most relevant here. Their overlay architecture \cite{Fang2017a} and identification of datacenters as an ideal place to perform such computations \cite{Huang2019a} matches our decisions. 
With respect to overlays, they implement blocks to accelerate the garbling of AND and XOR operations and that do not require the FPGA image to be recreated and programmed. Instead data is passed to these processing elements which is much more efficient.
We follow a similar scheme with Secret Sharing.

\section{Conclusion}
\label{sec:conclusion}

In this paper, we describe one approach to implementing the underlying MPC AND operation described by Araki et al. \cite{araki17} in hardware.
We demonstrate the viability of Secret Sharing MPC in a low latency environment and test the design on an FPGA in the cloud highlighting greater potential scalability of the design compared to alternatives.
With these insights, we plan to pursue improvements to this design to increase the performance further and to implement the higher level controls necessary to use our Secret Sharing building block in a complete MPC cloud service.


Some specific future work includes HDL implementation optimizations while maintaining the same scheme.
FPGA to FPGA communication will be evaluated. 
Additional research directions include different viable MPC security models and hardware security considerations on FPGAs.

\section{Acknowledgements}


Supported by Red Hat and NSF Grants 1718135, 1739000, and 1931714. DISTRIBUTION STATEMENT A. Approved for public release. Distribution is unlimited. This material is based upon work supported by the United States Air Force under Air Force Contract No. FA8702-15-D-0001. Any opinions, findings, conclusions or recommendations expressed in this material are those of the author(s) and do not necessarily reflect the views of the United States Air Force.


\bibliographystyle{IEEEtran}
\bibliography{ref/ref.bib}

\end{document}